\documentclass[aps,epsfig,article,11pt]{revtex4}

\usepackage{bm}
\def\bfvarphi{\mbox{\boldmath$\varphi$}}

\def\bfnabla{\mbox{\boldmath$\nabla$}}

\usepackage{latexsym,amsmath,tabularx,amssymb,bm}
\usepackage{graphicx}

\def\bra#1{\langle#1\vert}
\def\ket#1{\vert#1\rangle}

\newcommand{\pk}{\partial_\kappa}
\newcommand{\kp}{\kappa}
\newcommand \bea{\begin{eqnarray}}
\newcommand \eea{\end{eqnarray}}

\def\simge{\mathrel{%
   \rlap{\raise 0.511ex \hbox{$>$}}{\lower 0.511ex \hbox{$\sim$}}}}
\def\simle{\mathrel{
   \rlap{\raise 0.511ex \hbox{$<$}}{\lower 0.511ex \hbox{$\sim$}}}}

\def\simle{\mathrel{
   \rlap{\raise 0.511ex \hbox{$<$}}{\lower 0.511ex \hbox{$\sim$}}}}

\def\simge{\mathrel{%
    \rlap{\raise 0.511ex \hbox{$>$}}{\lower 0.511ex \hbox{$\sim$}}}}
\def\simle{\mathrel{
    \rlap{\raise 0.511ex \hbox{$<$}}{\lower 0.511ex \hbox{$\sim$}}}}
\newcommand{\beq}{\begin{eqnarray}}
\newcommand{\eeq}{\end{eqnarray}}
\newcommand{\del}{\partial}

\begin{document}

\title{Renormalization group flow equations
with full momentum dependence}

\author{Jean-Paul  Blaizot }

\affiliation{IPhT, CEA-Saclay, 91191 Gif-sur-Yvette, France}


\begin{abstract}
    After a short elementary introduction to the exact renormalization group for the effective action, I discuss a particular truncation of the hierarchy of flow equations that allows for the determination of the full momentum of the $n$-point functions. Applications are then briefly presented, to critical $O(N)$ models,  to Bose-Einstein condensation and  to finite temperature field theory. 
\end{abstract}

\maketitle


\section{Introduction}

\label{intro}

The exact, or non-perturbative, renormalization group (RG)  
\cite{Wetterich93,Ellwanger93,Tetradis94,Morris94,Morris94c}  stands
out as a very promising formalism to address non-perturbative
problems, i.e., problems whose solution does not appear to be expressible as an expansion in some small parameter. 
It leads to exact flow equations which cannot be solved in
general, but which offer the possibility for new approximation
schemes. 
It has been applied successfully to a variety
of problems, in condensed matter, particle or nuclear
physics (for  reviews with various points of view on the subject, see e.g.
\cite{Morris:1993qb,Bagnuls:2000ae,Berges02,Pawlowski:2005xe,Delamotte:2007pf,Rosten:2010vm}).

When only correlation functions at small momenta are
needed, as is the case for instance in the calculation of critical exponents, a general approximation method to solve the infinite
hierarchy of the flow equations has been developed
\cite{Morris94c,Bagnuls:2000ae,Berges02}. This method is based on a derivative expansion of  
the effective action.  However, in many
situations, this is not enough:  in order to calculate the
quantities of physical interest, the knowledge of the full momentum
dependence of the correlation functions is mandatory. 

The present paper deals with this issue, and summarizes work that has been done following the original suggestion by Blaizot, M\'endez-Galain and Wschebor (BMW)  \cite{Blaizot:2005xy,BMW-BE}  to obtain the momentum dependence of $n$-point functions from
the flow equations. The strategy put forward in
\cite{Blaizot:2005xy} is based on the fact that the RG flow at scale $\kappa$ involves the integration of fluctuations with momenta $q\leq \kappa$, as insured by the presence of a cutoff function $R_\kappa(q)$.   Since this cutoff function also guarantees that the vertex functions
are smooth functions of the momenta, these can be expanded in powers
of $q^2/\kappa^2$. The  ``leading order'' (LO) of the approximation scheme
 consists in a truncation at the level of the flow equation for the  two-point function $\Gamma_\kappa^{(2)}$,  setting $q=0$ in the
vertices $\Gamma_\kappa^{(3)}$ and $\Gamma_\kappa^{(4)}$ that appear in this flow equation. Doing so, and working in a constant external field, one can then express $\Gamma_\kappa^{(3)}$ and $\Gamma_\kappa^{(4)}$ as derivatives of $\Gamma_\kappa^{(2)}$ with respect to the background field, thereby closing  the hierarchy of flow 
equations. Next-to-leading orders are defined by similarly truncating the hierarchy at the level of higher point functions. 

The price to pay  is that the flow equations become 
differential equations with respect to a uniform background field,  with integral kernels that involve  the solution itself.
These non-linear integro-differential equations are a priori difficult to solve. It is possible to do so, however,  with a numerical effort comparable to that involved in solving the flow equations that result from the derivative expansion.  We shall provide here examples of results obtained with this method in various contexts.

The outline of the paper is as follows. In the Sect. \ref{method} we  briefly recall some basic features of the  exact renormalization group,   and derive  the flow equation for the effective action. We also digress on the use of the variational principle to obtain flow equations in the context of the hamiltonian formalism  often used in dealing with non relativistic systems. Then, in Sect. \ref{approx}, we discuss approximation schemes, such as the local potential approximation,  and  the BMW approximation scheme. Connections with the formalism of the two-particle irreducible effective action are also discussed. In Sect. \ref{applic}, we present three applications within scalar field theory with $O(N)$ symmetry in which  the full momentum dependence of the 2-point function plays an essential role. The paper ends with a short conclusion. 

\section{Exact renormalization group flow equations}
\label{method}

In this section we give an elementary introduction to the exact renormalization group, focussing on the effective action. For most of the discussion (except for the digression in subsection \ref{variation}) we shall restrict ourselves to the case of  a scalar field theory with
the classical action \beq\label{eactON} S = \int {\rm
d}^{d}x\,\left\lbrace{ \frac{1}{2}}   \left(\del_\mu
\varphi(x)\right)^2  + \frac{m^2}{2} \, \varphi^2(x) + \frac{u}{4!}
\,\varphi^4(x) \right\rbrace \,,\eeq
whose  parameters $m$ and $u$ are defined at some ``microscopic scale'' $\Lambda$, to be specified. 

\subsection{Flow equations for the $n$-point functions and their generating functional}

The strategy of the version of the renormalization group that we consider here is to build a family of theories indexed by a continuous parameter  $\kappa$, with the dimension of a momentum, and   
such that fluctuations are smoothly taken into account  as $\kappa$ is lowered 
from the  microscopic scale $\Lambda$ down  to 0. 
In practice, this is achieved by adding  to the original  
Euclidean action $S$ a (non-local)  term, quadratic in the fields, of the form   
\beq
\label{DeltaS}\Delta S_\kappa[\varphi]= \frac{1}{2} \int \frac{d^dq}{(2\pi)^d}
\,R_\kappa(q)\,\varphi(q)\varphi(-q).\eeq 
The regulator, or cut-off function, $R_\kappa(q^2)$ is chosen so that: i) $R_\kappa (q\ll \kappa)\sim \kappa^2$ which 
effectively suppresses the contribution of the modes $\varphi(q\lesssim  \kappa)$ by giving them a mass $\sim \kappa$;  ii)
it vanishes rapidly for $q\gtrsim \kappa$, leaving 
the modes $\varphi(q\gtrsim \kappa)$ unaffected. The explicit form of the cut-off function $R_\kappa(q)$ used in actual calculations will be specified  later. 

We consider then the  ``deformed'' (non local) field  theory with action $S_\kappa\equiv S+\Delta S_\kappa$, and the functional integral that yields the generating functional
of Green's functions 
 \beq Z_\kappa[j]=\int{\cal
D}\varphi \,{\rm e}^{-S_\kappa+\int j\varphi}, \eeq with $\int
j\varphi\equiv \int dx j(x)\varphi(x)$.
The expectation of the field
in the presence of the source $j(x)$ is given by 
\beq \label{phi}\frac{\delta
\ln Z_\kappa}{\delta j(x)}= \langle \varphi(x)\rangle=\phi(x).  \eeq 
Similarly
\beq\label{phiphi}
\partial_\kappa\ln Z_\kappa= -\frac{1}{2}\int_q\partial_\kappa R_\kappa(q) \langle
\varphi(q)\varphi(-q)\rangle, \eeq  
where we used here the fact that  all the
dependence on $\kappa$ is contained in the regulator term $\Delta S_\kappa[\varphi]$. Note that the expectation values in Eqs.~(\ref{phi}) and (\ref{phiphi}) are
taken in the presence of the regulator, and depend therefore on
$\kappa$. At this point, we define the effective action for the deformed theory,  $\Gamma_\kappa[\phi]$, 
through a Legendre transform \beq\label{Legendre} \Gamma_\kappa[ \phi ]+ \ln
Z_\kappa[j]=\int j \phi, \qquad\qquad \frac{d\Gamma_\kappa( \phi
)}{d \phi }=j. \eeq The functional $\Gamma_\kappa[\phi]$ is the generating functional of the one-line irreducible $n$-point functions of the deformed theory. Taking into account  that, at fixed $\phi$, $j$ depends on
$\kappa$, one easily obtains the flow of $\Gamma_\kappa[\phi]$  \beq
\partial \Gamma_\kappa[ \phi ]=\frac{1}{2} \int_q \partial_\kappa R_\kappa(q) \langle
\varphi(q)\varphi(-q)\rangle. \eeq It is in fact convenient to redefine $\Gamma_\kappa$ by subtracting from it 
$\Delta S_\kappa$. This subtraction has the advantage to make $\Gamma_\kappa$ coincide with the classical action $S$ at the microscopic scale $\Lambda$ (rather than to make it coincide with $S_\Lambda=S+\Delta S_\Lambda$, as would be the case for the definition (\ref{Legendre})). Furthermore, it modifies the flow equation in such a way that only the field fluctuations are involved. That is, once the subtraction is made, the flow is driven by the {\it connected} 2-point function \cite{Wetterich93}:
 \beq \label{NPRGeq}
\partial_\kappa \Gamma_\kappa[\phi]=\frac{1}{2} \int \frac{d^dq}{(2\pi)^d}
\,\partial_\kappa R_\kappa(q)\,
G_\kappa(q,-q;\phi),
\eeq
where 
\beq G(q, -q;\phi)=\langle \varphi(q)\varphi(-q)\rangle_c=\langle
\varphi(q)\varphi(-q)\rangle-  \phi(q) \phi(-q), \eeq
is the full propagator in the presence of the background field $\phi$. Using well-known properties of the Legendre transform, one can relate $G(q, -q;\phi)$ to $\Gamma^{(2)}_\kappa[\phi]$,  the second functional derivative of 
$\Gamma_\kappa[\phi]$ w.r.t. $\phi$:
\beq
G_\kappa^{-1}[\phi]=\Gamma_\kappa^{(2)}[\phi]+R_\kappa.
\eeq

The initial conditions on the flow equation 
 (\ref{NPRGeq}) are specified at 
the microscopic scale $\kappa=\Lambda$ mentioned above. This scale $\Lambda$ is the scale where fluctuations are damped by  
$\Delta S_\kappa$, so that $\Gamma_{\kappa=\Lambda}[\phi]\approx S[\phi]$. It should be emphasized however that this  issue of the initial condition at the scale $\Lambda$ is a subtle one. It is intimately related to that of renormalization and  ultraviolet divergences, issues that will not be discussed here.
The effective action of the original scalar field theory  is obtained as the solution of  Eq.~(\ref{NPRGeq})
for $\kappa\to 0$, at which point $R_\kappa(q^2)$ vanishes. Whether at that point the  result obtained for $\Gamma[\phi]$ is independent on the choice of $R_\kappa$, that is on the path joining the classical action to the full effective action, is another subtle question, whose answer depends on the approximation made. We shall comment on it further when we discuss specific approximations. 

By taking successive functional derivatives of  eq.~(\ref{NPRGeq}) with respect to $\phi$, and then
letting the field be constant, one gets the flow equation for the $n$-point functions 
\beq
\Gamma^{(n)}(x_1,\cdots ,x_n;\phi)\equiv \left. \frac{\delta^n\Gamma_\kappa}{\delta\phi(x_1) \dots
\delta\phi(x_n)}\right|_{\phi(x)\equiv \phi},
\eeq
 in a constant background field. Since the background is constant, these functions are invariant under translations of the coordinates, and it is convenient to factor out of the definition of their Fourier transform  the $\delta$-function that expresses the conservation of the total momentum. Thus, with an obvious abuse of notation,  we define the 
 $n$-point functions $\Gamma_\kappa^{(n)}(p_1,\dots,p_n;\phi)$ as: \beq\label{gamman} &&(2\pi)^d
\;\delta^{(d)}\left(p_1+\cdots
+p_n\right)\;\Gamma_\kappa^{(n)}(p_1,\dots,p_n;\phi)\qquad\qquad\nonumber
\\&& \qquad\qquad   \equiv \int d^dx_1\dots\int d^dx_{n} \,e^{i\sum_{j=1}^n
p_jx_j}\,\Gamma^{(n)}(x_1,\cdots ,x_n;\phi). \eeq 
We use here the convention of incoming momenta, and it is understood that in $\Gamma_\kappa^{(n)}(p_1,\dots,p_n;\phi)$ the sum of all momenta vanishes, so that $\Gamma_\kappa^{(n)}(p_1,\dots,p_n;\phi)$ is actually a function of $n-1$ momentum variables.  We shall often use the simplified notation $\Gamma_\kappa^{(2)}(p;\phi)$ for the function $\Gamma_\kappa^{(2)}(p,-p;\phi)$. 

When $\phi$ is constant, the functional $\Gamma_\kappa[\phi]$ itself reduces, to within a volume factor $\Omega$ to the effective potential $V_\kappa(\phi)$:
\beq
\Gamma_\kappa[\phi]=\Omega V_\kappa[\phi],\qquad \phi {\mbox  \; { \rm constant.}}
\eeq
The  flow equation for  the effective potential $V_\kappa$ follows from that of the effective action $\Gamma_\kappa$, Eq.~(\ref{NPRGeq}),  when restricted to constant $\phi$. It reads  \beq\label{eqforV} \kappa\partial_\kappa
V_\kappa(\phi)=\frac{1}{2}\int \frac{d^dq}{(2\pi)^d}\, \kappa
\partial_\kappa R_\kappa(q)\, G_\kappa(q,\phi), \eeq
where 
\begin{equation}\label{G-gamma2}
G^{-1}_{\kappa} (q,\phi) \equiv \Gamma^{(2)}_{\kappa} (q,\phi) +
R_\kappa(q).
\end{equation}

 By  taking two derivatives  of  eq.~(\ref{NPRGeq}) with respect to $\phi$, and then
letting the field be constant, one obtains the 
equation for the 2-point function:
\begin{eqnarray}
\label{gamma2champnonnul}
\partial_\kappa\Gamma_{\kappa}^{(2)}(p,\phi)&=&\int
\frac{d^dq}{(2\pi)^d}\,\partial_\kappa R_\kappa(q)\,G_{\kappa}^2(q,\phi)\nonumber\\
&\times&\left\{\Gamma_{\kappa}^{(3)}(p,q,-p-q;\phi) G_{\kappa}(q+p,\phi)\Gamma_{\kappa}^{(3)}(-p,p+q,-q;\phi)
-\frac{1}{2}\Gamma_{\kappa}^{(4)}
(p,-p,q,-q;\phi)\right\} .\nonumber \\\eeq
A diagrammatic illustration of this equation is given in Fig.~\ref{2-point-diagrams}.
\begin{figure}
\begin{center}
\includegraphics[scale=.4] {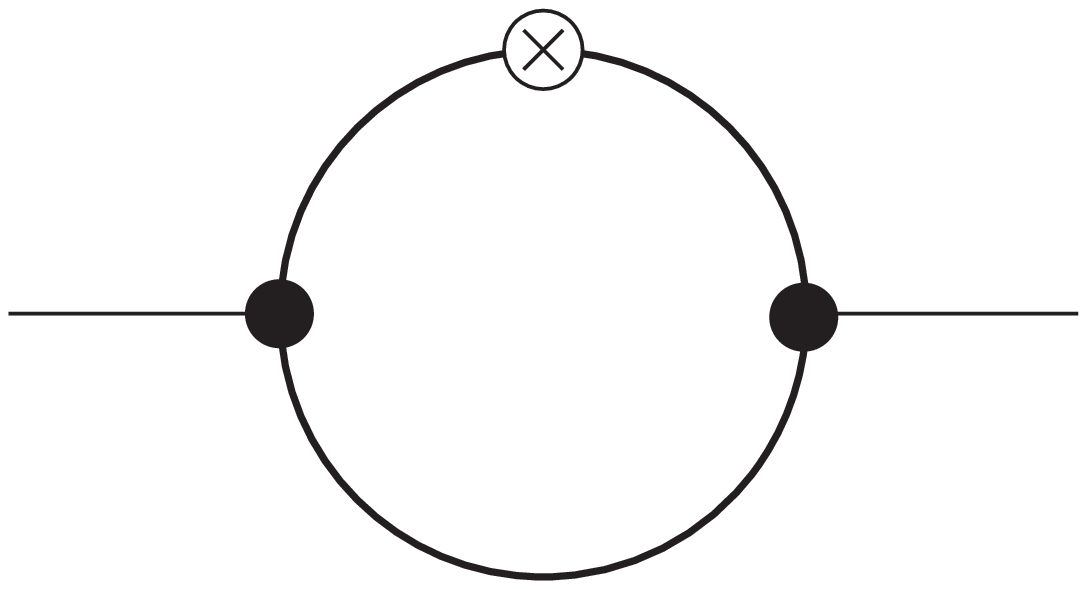} \hspace{20mm}
\includegraphics[scale=.4] {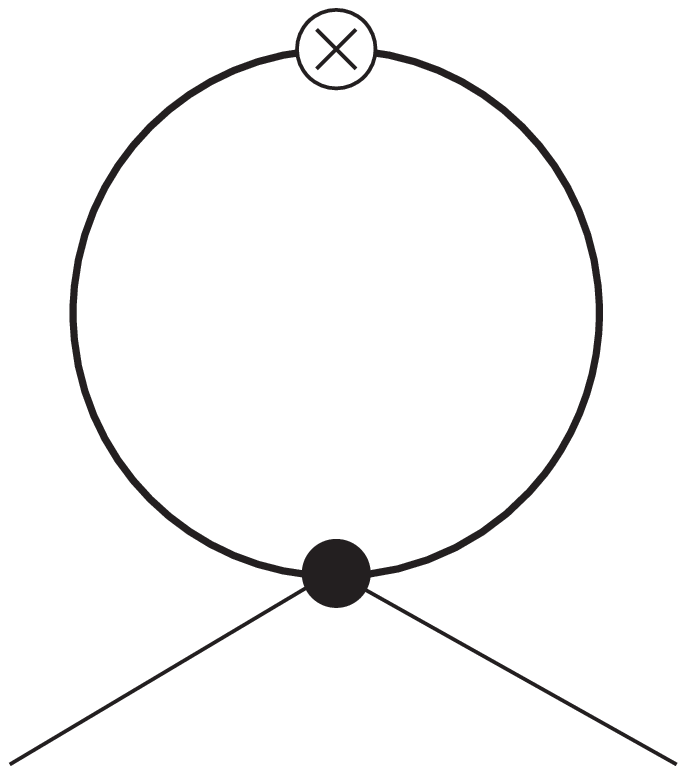}
\end{center}
\caption{The two diagrams contributing to the flow of the 2-point function, Eq.~(\ref{gamma2champnonnul}). The internal  lines represent dressed propagators, $G_\kappa$. The
 cross represents an insertion of $\partial_\kappa R_k$. The vertices denoted by   black dots are $\Gamma^{(3)}_\kappa$ (left) and $\Gamma^{(4)}_\kappa$ (right). \label{2-point-diagrams}}
\end{figure}
The flow equations for the $n$-point functions, such as Eq.~(\ref{gamma2champnonnul}), constitute  an infinite tower of coupled equations. The coupling between equations occurs in two ways, upwards and downwards. Upwards: typically the equation for $\Gamma^{(n)}$ involves  $\Gamma^{(n+1)}$ and $\Gamma^{(n+2)}$. Downwards:  all the flow equations involve $\Gamma^{(2)}$ which is coupled successively to all the equations above it.   There exist also consistency conditions that we illustrate here with $\Gamma^{(2)}$ and  the effective potential: $\Gamma^{(2)}(q=0)=\del^2 V/\del \phi^2$. Thus, $\Gamma^{(2)}(q=0)$ can be calculated by taking  the second derivative  with respect to $\phi$ of the flow equation for the effective potential:
\begin{eqnarray}
\label{gamma2champnonnul3}
\partial_\kappa\Gamma_{\kappa}^{(2)}(p=0,\phi)&=&\int
\frac{d^dq}{(2\pi)^d}\,\partial_\kappa R_\kappa(q)\,G_{\kappa}^2(q,\phi)\nonumber\\
&\times&\left\{G_{\kappa}(q,\phi)  \left(  \frac{\del \Gamma_{\kappa}^{(2)}(q;\phi)}{\del\phi} \right)^2
-\frac{1}{2}\frac{\del ^2\Gamma_{\kappa}^{(2)}
(q;\phi)}{\del\phi^2}\right\} .\nonumber \\\eeq
Alternatively, it can be obtained from Eq.~(\ref{gamma2champnonnul}) where we set $p=0$. The two calculations agree  owing to the fact that
\beq\label{derivs}
\Gamma_{\kappa}^{(3)}(q,-q,0;\phi)=\frac{\partial
\Gamma_{\kappa}^{(2)} (q;\phi)} {\partial \phi} , \hskip 1 cm
\Gamma_{\kappa}^{(4)}(q,-q,0,0;\phi)=\frac{\partial^2
\Gamma_{\kappa}^{(2)} (q;\phi)} {\partial \phi^2}. \eeq 
We  shall  consider later  approximation schemes that treat at different levels of accuracy the equations for various $n$-point functions. This may lead to violations of such consistency conditions if one is not careful. 

Relations such as (\ref{derivs}) between $n$-point functions with some vanishing external momenta and derivatives of $n$-point functions of lower rank with respect to a constant background field play an important role in the approximation scheme presented below.  In this context, we simply note here that the effective potential may be viewed as the generating functional of the $n$-point functions with vanishing external momenta. That is (for constant $\phi$)
\beq\label{generating0}
\Gamma_\kappa^{(n)}(0,\cdots , 0)=\frac{\del^n V_\kappa}{\del \phi ^n}.
\eeq
In the same way, we may regard $\Gamma_{\kappa}^{(2)}(p;\phi)$ as the generating functional of $n$-point functions with one non-vanishing momentum
\beq\label{generating2}
\Gamma_\kappa^{(n)}(p,-p,0,\cdots , 0)=\frac{\del^{n-2} \Gamma_{\kappa}^{(2)}(p;\phi)}{\del \phi ^{n-2}}.
\eeq

\subsection{Flow equation from the variational principle}
\label{variation}

The previous derivation of the flow equation did not refer to the interpretation of the various manipulations that were done in terms of ``coarse graining'' or ``elimination of degrees of freedom'', that are essential aspects of the renormalization group. These  are actually hidden in the specific functional form of the regulator $R_\kappa(q)$, but nothing in the derivation of the flow equation actually depends on the choice of the regulator (asides from being quadratic in the fields, but even that can be generalized). Thus, one may view the flow equation as merely  a tool to continuously go from a ``simple'' action (the classical action) to the full effective action, by tuning an appropriate contribution to the action. This point of view opens new perspectives and suggests that the strategy of flow equations may have a larger flexibility than is presently exploited. As an illustration, we shall here indicate how the variational principle can be used to establish flow equations in terms of hamiltonians and wave-functions for non-relativistic systems. 

The strategy we shall follow is identical to that of the previous section.  The system is described by a hamiltonian $H$, to which is added an
 ``external field''  $\Delta H_\kappa$ that  plays the role of the regulator. We assume that  $\Delta H_{\kappa=0}=0$, and that $H_\Lambda\equiv H+\Delta H_\Lambda$ is somehow ``simple'', so that  for instance one can determine its  ground state. Then the flow equation will take us from the ground state of this simple hamiltonian $H_\Lambda$ to the ground state of the hamiltonian $H$. 
 
 In this context, the initial condition of the flow is not restricted, as it is in most field theoretical applications to be some ``classical action''. Thus, it may be advantageous to depart from the standard approaches that would take $H_\Lambda$ to be the hamiltonian of independent particles, or quasi-particles,  and  include part of the effects of the interactions into $H_\Lambda$. In fact such a strategy was recently implemented by Machado and Dupuis, in their lattice renomalization group \cite{Machado:2010wi}, in which they separate the on-site physics form the long range correlations, only the latter being treated by the renormalization group. Much earlier, Parola and Reatto have developed a theory of liquids in which the short range part of the interaction is treated (almost) exactly with hard sphere systems, while the long range correlations are treated by a flow equation \cite {Reatto,Parola} that actually resembles closely Eq.~(\ref{NPRGeq}).
 
 The flow equation for the effective action, Eq.~(\ref{NPRGeq}), is general and  could be used in the non-relativistic many-body context. However, to bring a new perspective to the discussion, we shall go through an extremely simple derivation of a flow equation that relies on the 
variational principle \cite{Taeko}. For simplicity, we shall assume that $\Delta H_\kappa$ is a 
one body operator, i.e., a quadratic form of creation and destruction operators, that depends on the continuous parameter
$\kappa$:
 \beq
 \Delta H_\kappa=\sum_{\bf q} R_\kappa(q) a^\dagger_{\bf q} a_{\bf q},
 \eeq
 with $a^\dagger_{\bf q}$ and $ a_{\bf q}$ creation and destruction operators (of fermions or bosons). We wish to calculate the ground state energy  $\hat E_\kappa$ of the hamiltonian $H_\kappa = H+\Delta H_\kappa$. We call $\ket{\Psi_\kappa}$ the corresponding eigenstate. In fact we do not need the exact eigenstate, it is enough that $\ket{\Psi_\kappa}$ be determined from the variational principle
\beq
\left.\delta\bra{\Psi_\kappa} H+\Delta H_\kappa \ket{\Psi_\kappa}\right|_{R_\kappa}=0,
\eeq
where it is understood that the regulator $R_\kappa$ remains fixed in the variation.
The flow equation for  $\hat E_\kappa$ follows then immediately
 \beq
\partial_\kappa \hat E_\kappa= \sum_{\bf q}\partial_\kappa R_\kappa(q)\,
\bra{\Psi_\kappa} a^\dagger_{\bf q} a_{\bf q} \ket{\Psi_\kappa},  \eeq
where $\bra{\Psi_\kappa} a^\dagger_{\bf q} a_{\bf q} \ket{\Psi_\kappa}$ can be seen as an  occupation factor. The equation above is  the analog of Eq.~(\ref{phiphi}). It just describes the change in the ground state energy under the change of the parameters of the hamiltonian. Of course, when some creation or destruction operators acquire expectation values (as in the case of Bose-Einstein condensation), it may be convenient to work with the connected part of $\bra{\Psi_\kappa} a^\dagger_{\bf p} a_{\bf p} \ket{\Psi_\kappa}$, as we did earlier. Note that the flow of the energy $E_\kappa=\bra{\Psi_\kappa}H\ket{\Psi_\kappa}$ reads
\beq
\del_\kappa E_\kappa=-\sum_{\bf q} R_\kappa(q)\, \del_\kappa\bra{\Psi_\kappa} a^\dagger_q a_q \ket{\Psi_\kappa}.
\eeq

This approach is useful if we can determine (approximately) the wave function, or at least the occupation factors, as a function of $\kappa$. It was applied in the study of the BCS-BEC crossover \cite{Taeko}: in this case the BCS wave function leads to a reasonably accurate estimate of the occupation factors and their dependence on $\kappa$, and this extremely simple scheme allowed us to reproduce qualitatively (and even semi quantitatively) the results of more complete renormalization group studies, such as those carried out in Ref.~\cite{Birse:2004ha} .

\section{Approximation schemes}
\label{approx}

The flow  equations for the $n$-point functions that have been presented in the previous section are exact. Their solution requires, in general, approximations. It is precisely one of the virtues of the  formulation of field theory based on the exact renormalization group  to suggest approximations that are not easily derived in other, more conventional approaches.  Particularly interesting are the  approximation schemes  for the effective action itself, that is, approximations that  apply to  the whole set of $n$-point functions at once. The approximation schemes to the discussed in this section have this property. 

\subsection{The local potential approximation}

The local potential approximation (LPA) is the simplest of such approximations. It consists in assuming that for all values of $\kappa$ the effective action takes the form
\beq\label{GammaLPA}
\Gamma_\kappa[\phi]=\int d^dx\left\{ \frac{1}{2} \left( \del\phi\right)^2 +V_\kappa(\phi)\right\}.
\eeq
Within this approximation, the 2-point function, obtained by differentiating $\Gamma_\kappa[\phi]$ in Eq.~(\ref{GammaLPA}) twice with respect to $\phi$, and letting $\phi$ be constant,  is of the form ($\rho\equiv \phi^2/2$)
\beq\label{GLPA}
\Gamma^{(2)}_\kappa (q;\rho)=q^2+m_\kappa^2(\rho)\qquad m_\kappa^2(\rho)\equiv\frac{\partial^2
V_\kappa}{\partial \phi^2}.
\eeq
The corresponding propagator is simply a massive propagator, with a $\phi$-dependent mass $m_\kappa(\rho)$. The flow equation  for the effective potential $V_\kappa$  becomes a closed equation that we write as follows
 \beq\label{VeffI1}
\kappa\partial_\kappa
V_\kappa(\rho)=\frac{1}{2}I_1(\kappa,\rho),
\eeq
where $I_1(\kappa,\rho)$ is given by 
\beq\label{defI}
I_n (\kappa,\rho)\equiv J_n(p=0,\kappa,\rho),\qquad 
J_n(p,\kappa,\rho )\equiv
\int\frac{d^dq}{(2\pi)^d} \;\kappa\partial_\kappa R_\kappa(q)\;
G_\kappa(p+q )G^{n-1}_\kappa(q ). \eeq
 Note that the form (\ref{VeffI1}) of the equation for the effective potential is general: it would  yield the exact effective potential if the propagator used to calculate $I_1$ was the exact propagator, instead of the LPA propagator. Note also that, within the LPA,   the equation for the 2-point function at vanishing external momentum,  given by 
Eq.~(\ref{gamma2champnonnul3}),  reduces to a closed equation
\beq
 \label{2pointclosed0}
\kappa \partial_\kappa\Gamma_\kappa^{(2)}(0;\rho)=
I_3(\kappa,\rho) \; \left( \frac{\partial \Gamma_\kappa^{(2)}(0;\rho)} {\partial \phi}    \right)^2
  -\frac{1}{2} I_2(\kappa,\rho) \; \frac{\partial^2
\Gamma_\kappa^{(2)}(0;\rho)} {\partial \phi^2},
\eeq
since  the derivatives of $\Gamma_{\kappa}^{(2)}(q;\rho)$ with respect to $\phi$ are identical to derivatives of the effective potential, or, equivalently, to derivatives of $\Gamma_{\kappa}^{(2)}(0;\rho)$.

The LPA has been widely used, and given its simplicity, the quality of the results obtained is quite good  \cite{Berges02,delamotte03,canet04b,canet05a}. Note that the LPA, by construction, yields $n$-point functions that have no intrinsic dependence on the external momenta: the generating functional of these $n$-point functions is given by  Eq.~(\ref{GammaLPA}) taken at $\kappa=0$, so that, for $n>2$, the LPA $n$-point functions are just derivatives of the effective potential (see Eq.~(\ref{generating0})). 
The LPA can be improved through an expansion in gradients of the fields, 
usually referred to as  the derivative expansion (DE) \cite{Morris:1993qb,Berges02,canet03a}. 
Within the derivative expansion, the lowest $n$-point functions  have polynomial dependence on the momenta (the degree of the polynomial corresponding to the order of the expansion in the derivatives). It follows in particular that the derivative expansion does not directly describe the  anomalous behavior of $n$-point functions at small momenta (anomalous dimensions can be recovered for instance from the scale dependence of the field normalization). In order to capture the full momentum dependence of the $n$-point function, a better truncation scheme is necessary:  the BMW scheme \cite{Blaizot:2005xy}, to be discussed next, achieves this goal.

\subsection{The approximation BMW-LO}

The  BMW approximation scheme  
 relies on two observations. 
First, the presence of the cut-off function $R_\kappa(q)$  insures
that the $n$-point functions $\Gamma^{(n)}_\kappa(p_i)$ remain regular functions of the external momenta $p_i$ as $p_i\to 0$; besides,  it limits  the internal momentum $q$ in equations such as Eq.~(\ref{gamma2champnonnul})
to $q\lesssim \kappa$. In line with this observation, the approximation consists in neglecting the $q$-dependence of the vertex functions in the r.h.s. 
of the flow equations (e.g. set $q=0$ in $\Gamma^{(3)}$  and $\Gamma^{(4)}$  in Eq.~(\ref{gamma2champnonnul})),  
while keeping the full dependence on the external momenta $p_i$.  
The second observation is that, for uniform fields,
$\Gamma^{(m+1)}_k(p_1,\dots,p_m,0,\phi)=
\partial_\phi \Gamma^{(m)}_k(p_1,\dots,p_m,\phi)$ (a relation that we have already used earlier, see Eq.~(\ref{derivs})). This  enables one to close the hierarchy of equations at some finite order.  The order $m$ of the scheme consists in keeping the full momentum 
dependence of  all the $n$-point functions up to $\Gamma^{(m)}_\kappa$, and expressing   
$\Gamma^{(m+1)}_\kappa$ and $ \Gamma^{(m+2)}_\kappa$ as derivatives of $\Gamma^{(m)}_\kappa$ with respect to $\phi$, after setting to zero the loop momenta that flows through $\Gamma^{(m+1)}_\kappa$ and $ \Gamma^{(m+2)}_\kappa$ in the equation for $\Gamma^{(m)}_\kappa$. The accuracy of the scheme depends of course of the rank $m$  at which one operates the truncation, but obviously the implementation becomes increasingly complicated as $m$ grows. It is therefore gratifying that accurate results can be obtained with lowest order truncations.

The approximations obtained by truncating the hierarchy at the lowest level, i.e., $m=0$ is identical to the local potential approximation discussed in the previous subsection. The next order of the approximation scheme,  which we shall in fact refer to  as the leading order (LO) of the BMW method, consists in a truncation at the level $m=2$, i.e., at the level of the flow equation of the 2-point function. The resulting equation is the closed equation
 \beq
 \label{2pointcloseda}
\kappa \partial_\kappa\Gamma_\kappa^{(2)}(p,\rho)=
J_3(p,\kappa,\rho) \; \left( \frac{\partial
\Gamma_\kappa^{(2)}(p,\rho)} {\partial \phi} \right)^2
  -\frac{1}{2} I_2(\kappa,\rho) \; \frac{\partial^2
\Gamma_\kappa^{(2)}(p,\rho)} {\partial \phi^2},
\eeq
where $J_3(p,\kappa,\rho)$ and $I_2(\kappa,\rho)$ are obtained from the general definitions in Eq.~(\ref{defI}). 
This equation may be viewed as a generalization of Eq.~(\ref{2pointclosed0}) that takes into account the full momentum dependence of the 2-point function. What the BMW approximation achieves is the possibility to factorize the vertices and take them out of the integrals. The momentum dependence that remains within the 3 and 4-point vertices is that of the 2-point function itself (see Eq.~(\ref{generating2})). Interestingly, a very similar set of equations  have been obtained much earlier in the context of the theory of liquids \cite{Reatto}, but these were unknown to the authors of Ref.~\cite{Blaizot:2005xy}.

At this point we recall the consistency condition discussed in connection with Eq.~(\ref{gamma2champnonnul3}): the flow equation  for $\Gamma^{(2)}_\kappa
(p=0,\rho)$ can be obtained by taking the second  derivative of the flow equation for the effective potential  with respect to the
background field. The resulting equation,   Eq.~(\ref{gamma2champnonnul3}),  does not coincide here with Eq.~(\ref{2pointcloseda})
in which we set $p=0$. This is because  Eq.~(\ref{2pointcloseda}) results from the BMW truncation, which is not implemented in Eq.~(\ref{gamma2champnonnul3}), and needs not be. This results in a mismatch between the two ways of calculating $\Gamma^{(2)}(p=0)$, whose origin can be traced back to the fact that the 2-point function and the effective potential are not determined with the same accuracy: loosely speaking, the 2-point function calculated from Eq.~(\ref{2pointcloseda}) is accurate to order one-loop, while the effective potential obtained from Eq.~(\ref{VeffI1}), with the BMW propagator, is accurate to order two-loop. In order to properly deal with this feature, we treat separately the zero momentum ($p=0$) and the non-zero momentum
($p\ne 0$) sectors, and  
 write 
\beq\label{defsigmabis}
\Gamma^{(2)}_\kappa (p,\rho)\equiv  p^2+ \Delta_\kappa (p,\rho)+m_\kappa^2(\rho),
\eeq
where $ \Delta_\kappa (p=0,\rho)=0$.
Now, $m_\kappa^2(\rho)$ is obtained by solving the equation for the effective potential, while the equation for $\Delta_\kappa(p,\rho)$ can be easily deduced from that for $\Gamma^{(2)}_\kappa$, i.e., from Eq.~(\ref{2pointcloseda}) by subtracting the corresponding equation that holds for $p=0$. It reads
\beq\label{2pointclosedab0}
 \partial_t\Delta_\kappa(p,\rho)&=&
2\rho  J_3(p,\kappa,\rho) \; \left[u_\kappa(\rho)+\Delta_\kappa^\prime(p,\rho)\right]^2-2\rho I_3(\kappa,\rho)\; u_\kappa^2(\rho)  \nonumber\\
  &-&\frac{1}{2} I_2(\kappa,\rho) \; \left[
\Delta_\kappa^\prime(p,\rho)+2\rho
\Delta_\kappa^{\prime\prime}(p,\rho)\right],
\eeq
where  the symbol $^\prime$ denotes the derivative with respect to $\rho$, and $u_\kappa(\rho)\equiv dm_\kappa^2(\rho)/d\rho$.

This equation (\ref{2pointclosedab0}), together with that for the effective potential, Eq.~(\ref{VeffI1}), 
and that for the propagator
\beq
G_\kappa^{-1}(q,\rho)=q^2+\Delta_\kappa(q,\rho)+m_\kappa^2(\rho)+R_\kappa(q),\eeq
constitute a closed system of equations for $\Gamma_{\kappa}^{(2)}(p,\rho)$ which can be solved with the initial condition $\Gamma_{\Lambda}^{(2)}(p,\rho)=p^2+m^2+u \rho$.

 \subsection{Relation to the 2PI formalism}
 
 Because the exact RG formalism that we are using puts emphasis on the propagator, it is natural to ask for the connection with the 2PI (2-particle irreducible) formalism \cite{LW,Baym,Cornwall:vz}.  Let us   recall that the central quantity in this formalism  is $\Phi[G]$, the sum of the two-particle-irreducible ``skeleton'' diagrams, a  functional of the full propagator $G$. From $\Phi[G]$  one obtains the self-energy by functional differentiation (to within factors $(2\pi)^3)$:
\beq\label{PhiPi}
\Sigma(p)=2 \frac{\delta \Phi}{\delta G(p)}.
\eeq
This relation, together with Dyson's equation:
\beq\label{Dyson}
G^{-1}(p)=p^2+m^2+\Sigma(p)\,,
\eeq
defines the physical propagator and self-energy in a self-consistent way. We shall refer to Eq.~(\ref{Dyson}), with $\Sigma[G]$ given by Eq.~(\ref{PhiPi}), as the ``gap equation''. A further differentiation of $\Phi[G]$ with respect to $G$ yields the two-particle-irreducible kernel
\beq\label{eq:Lambda}
{\cal I}(q,p)=2\frac{\delta \Sigma(p)}{\delta G(q)}=4\frac{\delta^2\Phi}{\delta G(q)\delta G(p)}={\cal I}(p,q),
\eeq
of a Bethe-Salpeter  type equation
\beq\label{BS1}
\Gamma^{(4)}(q,p) & = & {\cal I}(q,p)-\frac{1}{2}\int_l \Gamma^{(4)}(q,l)\,G^2(l)\,{\cal I}(l,p),
\eeq
that allows the calculation of the four-point function $\Gamma^{(4)}(q,p)\equiv\Gamma^{(4)}(q,-q,p,-p)$: the quantity ${\cal I}(q,p)$ is the two-particle-irreducible contribution to $\Gamma^{(4)}(q,p)$ in one particular channel. If all skeletons are kept in $\Phi$, these relations  are exact. A $\Phi$-derivable
approximation \cite{Baym} is  obtained by selecting a class of skeletons in $\Phi$ and calculating $\Sigma$ and $\Gamma^{(4)}$ from the equations above.

The 2PI formalism provides a set of  functional relations among the $n$-point functions that can be used to define a truncation of the flow equations \cite{Blaizot:2010zx}. Consider indeed the equation for the 2-point function
for a vanishing background field. It reads:
\beq
\pk \Gamma^{(2)}_\kappa(p)= & - & \frac{1}{2}\int_q \,\pk R_\kappa(q)\,G^2_\kp(q)\,\Gamma^{(4)}_\kp(q,p)\label{eq:flow2}.\eeq
A possible truncation consists in using for $\Gamma^{(4)}_\kp(q,p)$ in the right-hand-side of this equation, the extension of the relation (\ref{BS1}) to the deformed theory, namely 
\beq
\Gamma^{(4)}_\kp(q,p)={\cal I}_\kp(q,p) & - & \frac{1}{2}\int_l\,\Gamma^{(4)}_\kp(q,l)\,G^2_\kp(l)\,{\cal I}_\kappa(l,p)\,,\label{eq:BS2}
\eeq
where the subscript $\kappa$ on $\mathcal{I}_\kappa$ means that the functional derivative defining the kernel $\mathcal{I}$ (see Eq.~(\ref{eq:Lambda})) is to be evaluated for $\smash{G=G_\kappa}$, with $G_\kappa^{-1}=\Gamma^{(2)}_\kappa+R_\kappa$. Since $\mathcal{I}_\kappa$ is a functional of the 2-point function, the system of equations (\ref{eq:flow2}-\ref{eq:BS2})  is indeed closed. 

One nice feature of this truncation scheme is that it is systematically improvable, by adding more skeletons to $\Phi$: if all skeletons are included, the solution of the coupled system of equations (\ref{eq:flow2}-\ref{eq:BS2}) provides the exact 2-point function as well as the exact 4-point function for a particular configuration of the external momenta. A second attractive feature is that it preserves the property of the flow of being a total derivative with respect to the parameter $\kappa$. It is indeed not difficult to show, using  Eqs.~(\ref{eq:BS2}) and  (\ref{eq:flow2}), that
\beq\label{eq:exact}
\pk\Gamma^{(2)}_\kp(p)=\pk\Sigma_\kappa(p),
\eeq
where  $\Sigma_\kappa\equiv \Sigma[G_\kappa]$, with $\Sigma[G]$ given by Eq.~(\ref{PhiPi}). This is a unique property of this truncation, that is not shared by most other popular truncations of the exact RG (with the noticeable exceptions of the perturbative expansion, and the large $N$ approximation, see e.g. \cite{Blaizot:2008xx}. A similar property of the flow equation was also obtained in the off-equilibrium context in Ref.~\cite{Gasenzer:2008zz}. The BMW truncation does not respect this property).  Since it is a total derivative, the flow can be easily integrated to yield the gap equation whose solution is equivalent to a resummation of  all the  Feynman diagrams  that are generated from the  skeletons that are kept in the approximation considered.  And because the solution of the gap equation corresponds to an exact resummation of selected Feynman diagrams, at the end of the flow where $\kappa=0$ and the regulator vanishes, the final result is rigorously independent of the choice of the regulator. 

All these properties of the 2PI truncation may look at first disappointing from the point of view of the flow equations: indeed, all what the flow does in this particular truncation is solving the 2PI equations! However, there is certainly interest in establishing direct connections between non trivial non-perturbative approximations. In particular, because the 2PI truncations lead to flow equations that are exact derivatives, they could be used to test other approximations. Besides, from the point of view of the 2PI formalism, there is a practical advantage in reformulating the gap equation as a flow equation:  this is because initial value problems are in general easier to solve than non linear gap equations. Furthermore, regarding the 2PI equations as flow equations shed a new light on the renormalization of $\Phi$-derivable approximations \cite{vanHees:2001ik,BPR}.

\section{Applications}\label{applic}

We turn now to a few   specific applications of the non perturbative renormalization group using the BMW truncation scheme. Most of these applications concern scalar field theories with an $O(N)$ symmetric  action of the generic form
\beq\label{eactONv} S = \int {\rm
d}^{d}x\,\left\lbrace{ \frac{1}{2}}   \left(\del_\mu
\bfvarphi(x)\right)^2  + \frac{m^2}{2} \, \bfvarphi^2(x) + \frac{u}{4!}
\,(\bfvarphi^2(x) )^2\right\rbrace, \qquad \bfvarphi^2=\sum_{a=1}^N \varphi_a\varphi_a.\eeq
The first application covered in this section may be seen as a ``classic'' application within the context of the renormalization group: it concerns the critical $O(N)$ models, and the calculation of critical exponents and the scaling functions \cite{Benitez:2009xg,Benitez09}. The second application refers to the shift of the transition temperature of the Bose-Einstein condensation in a dilute Bose gas: this again involve the $O(N)$ field theory (with $N=2$), but the relevant quantity to be calculated is sensitive to the whole momentum range of the 2-point function, and not only to the critical momentum region \cite{Blaizot:2008xx}. The last application concerns the thermodynamics of quantum fields. It shows how the exact renormalization group allows one to circumvent the specific difficulties of perturbation theory in such systems. The physics motivation is the physics of the quark-gluon plasma, but some of the essential difficulties of perturbation theory in QCD at finite temperature are well illustrated again by the scalar field \cite{Blaizot:2006rj,Blaizot:2010ut}. 

Before turning to these applications, we need to  comment on the dependence of the regulator in the practical calculations. We shall use in our calculations a   
 cut-off function of the generic form 
\beq \label{regulator}
R_\kappa(q)=Z_\kappa \kappa^2 r(\tilde q),\qquad \tilde q\equiv\frac{q}{\kappa}, 
\eeq
where $Z_\kappa$ is a function of $\kappa$ only. This factor $Z_\kappa$ reflects the finite change in normalization of the field between the scale $\Lambda$ and the scale $\kappa$. It can be defined as 
\begin{equation}\label{Zkapparho}
Z_\kappa=\left.\frac{\partial \Gamma^{(2)}_{\kappa}(p;\rho)}{\partial p^2}\right|_{p_0; \rho_0},
\end{equation}
with $\rho_0$ and $p_0$ a priori arbitrary. In practice, one usually chooses $p_0=0$ and $\rho_0$ the value of $\rho$ at the minimum of the effective potential. The factor $Z_\kappa$ enters the scaling to dimensionless variables used for the numerical solution in the critical region. 

The results of the calculations presented below were obtained with an exponential regulator
\beq
r(\tilde q)=\frac{\alpha \tilde q^2}{{\rm e}^{\tilde q^2}-1},\eeq
where $\alpha$ is a free parameter.
As we shall see, physical quantities exhibit a small
dependence  on  $\alpha$.  
Since in the absence of any approximation, they would be strictly independent of the 
cut-off function, 
 a study of this spurious dependence  provides an indication of the quality of the 
 approximation \cite{Litim:2001dt,Pawlowski:2005xe}.

\subsection{Critical $O(N)$ models}

The equations of the BMW method have been solved first with additional approximations in Refs.~\cite{BMWnum}, but the results presented here were obtained by solving numerically the nonlinear integro-partial-differential 
equations  (\ref{2pointclosedab0}) and (\ref{VeffI1}) without any further approximation \cite{Benitez:2009xg}. The numerical techniques used are described in  Refs.~\cite{Benitez:2009xg} and \cite{Benitez09}. Here, we report only some selected results, and comment on important aspects of the approximation that are needed in order to gauge the quality of these results.

Table \ref{tableexpo} contains 
results for  the critical exponents  $\eta$ and $\nu$, in dimension $d=3$ and for various values of $N$, together with some of the best estimates available in the literature. To appreciate the quality of these results, let us recall that these depend a priori on the parameters $\alpha$, ${p}_0$, and ${\rho}_0$ (see Eqs.~(\ref{regulator}) and (\ref{Zkapparho})). (They also depend on the functional form of the cut-off function, but no systematic exploration has been performed to study this dependence.)
In all cases studied,
we find the dependence on ${p}_0$ and $\rho_0$ to be much smaller than that
on $\alpha$, so that only the latter needs to  be considered. 
As a function of $\alpha$, 
physical quantities typically exhibit 
a single extremum  $\alpha^*$ located near $\alpha=2$. 
Moreover, we find this extremum generally pointing towards the best numerical estimates. Since, in the absence of any approximation there should be no dependence on $\alpha$, 
we regard these extremum values, being locally independent of $\alpha$, 
as our best values, adopting in doing so a strategy  often referred to as the principle of minimal sensitivity (PMS) \cite{Stevenson}. Our numbers are then all given for the PMS values $\alpha^*$ of the regulator parameter, 
and the digits quoted are those which remain stable  when 
$\alpha$ varies in the  range $[\alpha^*-\frac{1}{2},\alpha^*+\frac{1}{2}] $.
The quality of the values obtained for the critical exponents is obvious: 
For all $N$  the  
results for $\nu$ agree with the best  
estimates to within less than a percent; as for the values of  $\eta$ it is 
typically at the same distance from the 
Monte-Carlo and temperature series estimates as resummed perturbative data. 
For $N=100$, we find  $\eta=0.0023$, and $\nu=0.990$, which compare well to the 
values $\eta=0.0027$ and $\nu=0.989$ obtained in the $1/N$ expansion \cite{Moshe:2003xn}.
Our numbers also compare favorably with those obtained at order 
$\partial^2$ in the DE scheme\cite{canet03a}.

\begin{table}[tp]
\caption{\label{tableexpo} Critical exponents for the $O(N)$ models for $d=3$.}
\begin{ruledtabular}
\begin{tabular}{cllcllllcllll}
$N$& \multicolumn{2}{c}{BMW} &\ \  & \multicolumn{3}{c}{Resummed pert. exp.}&\ \ &\multicolumn{3}{c}{MC and high-temp. series} \\
   & \multicolumn{1}{c}{$\eta$} & \multicolumn{1}{c}{$\nu$}  &
   & \multicolumn{1}{c}{$\eta$} & \multicolumn{1}{c}{$\nu$}   & \multicolumn{1}{c}{Ref.} &
   & \multicolumn{1}{c}{$\eta$} & \multicolumn{1}{c}{$\nu$} &    
\multicolumn{1}{c}{Ref.} \\ \hline
0 & 0.034 & 0.589  & & 0.0284(25) & 0.5882(11) &   \cite{Guida98} &
   &0.030(3) & 0.5872(5) &   \cite{grassberger} \\
1  & 0.039 & 0.632&  & 0.0335(25) & 0.6304(13) &   \cite{Guida98} &
 & 0.0368(2) & 0.6302(1) & \cite{Deng03} \\
2 & 0.041 & 0.674 & & 0.0354(25) & 0.6703(15) &  \cite{Guida98}&
   & 0.0381(2)& 0.6717(1) &  \cite{Campostrini06}\\
3  & 0.040 & 0.715&  & 0.0355(25) & 0.7073(35) &  \cite{Guida98} &
  & 0.0375(5)& 0.7112(5) & \cite{Campostrini01}\cite{Hasenbusch01} \\
4 & 0.038 & 0.754&  & 0.035(4)& 0.741(6)  &   \cite{Guida98}&
    & 0.0365(10) & 0.749(2) & \cite{Hasenbusch01}\\
10&0.022 & 0.889& & 0.024 & 0.859 &    \cite{Antonenko98} &&&&&&\\
\end{tabular}
\end{ruledtabular}
\end{table}

The two-dimensional case, for which exact results exist,  
provides an even more stringent test of the BMW scheme. We have results at the moment only for the Ising 
model, i.e., for $N=1$, which exhibits  a standard 
critical behavior  in $d=2$. The perturbative method that works well in $d=3$ fails here:
for instance,  the fixed-dimension expansion that provides the best results in 
$d=3$, yields, in $d=2$ and at five loops, $\eta=0.145(14)$ \cite{pogorelov07}, in  
contradiction with the exact value $\eta=\frac{1}{4}$.  
We find instead $\eta=0.254$, $\nu=1.00$. Note however that no systematic study of the dependence of this result on the regulator parameter has been performed yet. 

The  BMW scheme  yields the complete momentum dependence of the 2-point function. All expected features of $\Gamma^{(2)}_k(p)$ at criticality 
are observed: 
In the infrared (IR) regime $\kp\ll p \ll u$, $\Gamma^{(2)}_\kp(p,0)\sim p^{2-\eta }\kp^\eta$, and 
this IR behavior of 
$\Gamma^{(2)}_\kp(p)$ can be used to extract the value of  $\eta$; 
the value obtained directly from the momentum dependence of $\Gamma^{(2)}_\kp(p)$ is in excellent agreement with that deduced  from the $\kappa$-dependence of the field normalization factor $Z_\kappa$. The ultraviolet (UV) regime $\kp,u\ll p \ll \Lambda$ exists 
if $u$ is sufficiently small; this regime  
can be studied perturbatively and one finds that, in leading order, $\Gamma^{(2)}_\kp(p,0))\sim u^2 \log (p/u)$. 
The present approximation reproduces this logarithmic behavior with, however, 
a prefactor 8\% larger than the  two-loop result. Note that the complete two-loop behavior can be 
recovered by a simple improvement of the BMW scheme \cite{Benitez09}.

\begin{figure}
\hfill{}\includegraphics[angle=-90,scale=0.4]{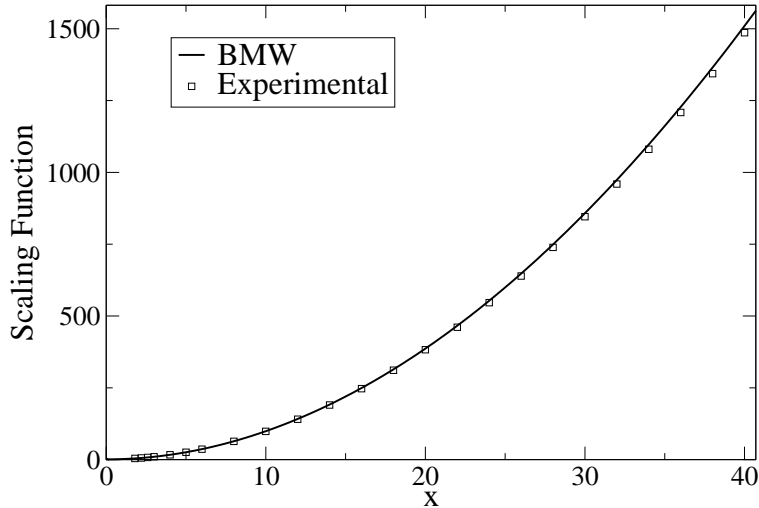}\hfill{}
\caption{\label{fig:scaling}The scaling function $g_+(x)$ as a function of $x=q\xi$. The experimental data are from Ref.~\cite{damay98}.}
\end{figure}

The momentum dependence of the  2-point function can be further tested by analyzing the so-called scaling functions
\beq
G_\pm (q)=\chi g_\pm(q\xi),
\eeq
where $\chi^{-1}\equiv\Gamma^{(2)}(q=0)$, and $\xi$ is the correlation length, which diverges close to criticality with the $\nu$ critical exponent. Here $\pm$ refers to the two phases, above and below the critical point, respectively. The functions $g_\pm$ are universal. We consider here the case $N=1$. The scaling function $g_+(x=q\xi) $ has been calculated in the BMW approximation for different values of the correlation length. It  is plotted in Fig.~\ref{fig:scaling}, where we can see that the scaling is perfectly reproduced. 
The agreement with the experimental data is also excellent, although this is somewhat delusive since, in this range of momenta, the physics is dominated by mean field effects, and the scaling function deviates only slightly from that predicted by the Ornstein-Zernicke approximation). A critical study is presented in Ref.~\cite{Benitez09}.

\subsection{Temperature of Bose-Einstein condensation}

A quantity particularly sensitive to the  UV-IR crossover region is  the shift, due to interactions,  of the critical temperature of the dilute, weakly interacting,  Bose gas \cite{Baym99}. The hamiltonian describing such a system is typically of the form
\beq \label{Hamiltonian}H=\int{d^3 {\bf r}} \left\{
 \psi^\dagger({\bf r}) \left(\-\frac{\bfnabla^2}{2m}\,
\!-\!\mu \right) \psi ({\bf r})+\frac{g}{2}
\psi^\dagger({\bf r})
\psi^\dagger ({\bf r})
\psi({\bf r})\psi({\bf r})
\right\},
\end{eqnarray}
where $g= 4\pi a/m$, with $a$ the s-wave scattering length.  (We are ignoring here a subtlety related to the ultraviolet divergences that are generated in calculations with a contact interaction and that require the introduction of an ultraviolet cut-off. This can be implemented in a standard fashion, but it plays no role in the present discussion.) 
The effective hamiltonian (\ref{Hamiltonian}) provides an accurate description of   dilute systems, when the scattering length is small compared to the interparticle distance, i.e.,   $an^{1/3}\ll 1$.  In the vicinity of the condensation, where $n^{1/3}\lambda\simeq 1$, with $\lambda=\sqrt{\frac{2\pi}{mT}} $  the thermal wavelength,  the diluteness condition reads $a/\lambda\ll 1$.

 It has been shown that the shift $\Delta T_c$ of the Bose-Einstein condensation temperature 
is linear in  $an^{1/3}$ \cite{Baym99}:
\beq\label{deltaTc}
\frac{\Delta T_c}{T_c^0}=\frac{T_c-T_c^0}{T_c^0}=c \,\,a n^{1/3}.
\eeq
Here $T_c^0$ is the
condensation temperature of the ideal gas, defined by the condition $n\lambda^3= \zeta(3/2)\approx 2.612$, and $T_c$ the
transition temperature of the interacting system at the same density.  This  result is non trivial: although the shift is proportional to $a$, and hence is small if $a$ is small, the result (\ref{deltaTc}) cannot be obtained form perturbation theory. It  is  obtained through the following chain of arguments. In the limit of small coupling the shift of the critical temperature may be obtained from the shift of the critical density, $\frac{\Delta T_c}{T_c^0}=-\frac{2}{3}\frac{\Delta n_c}{n_c^0}$. The latter quantity is easier to calculate, and is dominated by the contribution of the zero Matsubara frequency part, $\psi_0$, of the bosonic field, whose dynamics is governed by a three-dimensional classical field theory with $O(2)$ symmetry. To make the connection with the notation used in the rest of this section, we set $\psi_0=\sqrt{mT}(\varphi_1+i\varphi_2)$, with $\varphi_1$ and  $\varphi_2$ two real fields.  The action for these real fields is  given by Eq.~(\ref{eactONv}), with the parameters $r$ and $u$ related to the  parameters of the hamiltonian (\ref{Hamiltonian}) by
\beq\label{def:rmu}
r=-2mT\mu, \qquad u=96 \pi^2\frac{a}{\lambda^2}.
\eeq
The resulting density shift is then given by the change in the fluctuation of   this classical field, with the coefficient $c$ in Eq.~(\ref{deltaTc}) given by
\beq
c\,=-\frac{256 \pi^3}{\left(\zeta(3/2)\right)^{4/3}} \,
\frac{\Delta\langle\varphi_i^2\rangle}{Nu},
\eeq
in the limit  $u\to 0$ (and for $N=2$). 

The calculation of $\Delta\langle\varphi_i^2\rangle$ is difficult for the reason already mentioned: although the coupling constant $u$ can be arbitrarily small, perturbation theory cannot be used because of the infrared divergences of the critical three-dimensional field theory. The best numerical estimates for $\Delta\langle \varphi^2\rangle$,
and hence for $c$,  are those which have been obtained using the
lattice technique by two groups, with the results: $c=1.32\pm0.02$
\cite{latt2} and $c=1.29\pm 0.05$ \cite{latt1}. The
availability of these results has turned the calculation of $c$
into a testing ground for other non perturbative methods:
expansion in  $1/N$ \cite{Baym00,Arnold:2000ef}, optimized
perturbation theory \cite{souza,Kneur04},
   resummed    perturbative  calculations to high loop orders
\cite{Kastening03}. 

To understand better the origin of the difficulty of the calculation of $c$, as well as the linearity in $a$ of $\Delta T_c$, let us write
$\Delta\langle \varphi_i^2\rangle$ as the following integral
\beq\label{integralc}
\frac{\Delta\langle \varphi_i^2\rangle}{N}=
\int\frac{d^3 p}{(2\pi)^3}\,\left(\frac{1}{p^2+\Sigma(p)}-\frac{1}{p^2}\right)= -\frac{1}{2\pi^2}\int\frac{dp}{p}\left[ p-\frac{p^3}{p^2+\Sigma(p)}\right].
\eeq
where
$\Sigma(p)$ is the self-energy at  criticality, i.e.,
$\Sigma(0)=0$. In eq.~(\ref{integralc}),  the term within the square brackets is, to a very good  approximation (when  $p/u$ is small enough), equal to $\Sigma(p)/p$, a function that is peaked in the region of intermediate momenta between
the  critical region  and  the
high momentum perturbative region.
Thus, the difficulty in getting a precise evaluation of the integral
(\ref{integralc}) is that it requires an  accurate determination
of $\Sigma(p)$ in a large region of momenta including the
crossover region between two different physical regimes
\cite{bigbec,Baym00}. In that sense, the calculation of $c$ can be viewed as a very stringent test of the approximation scheme, and in fact, it is in order to obtain a reliable estimate of $c$ that the BMW scheme was initially developed.

To see the origin of the linear relation between $\Delta n_c$ and $a$, we note first that the action (\ref{eactON}) contains a single dimensionfull parameter, $u$, $r$ being adjusted for any given $u$ to be at criticality. In fact the effective three dimensional theory is ultraviolet divergent, so there is a priori another parameter, an ultraviolet cut-off $\Lambda\sim 1/\lambda$, with $\lambda$ the thermal wavelength. It follows then from dimensional analysis  that $\Sigma(p)$ can be written as
$
\Sigma(p=xu)=u^2 \sigma(x,u/\Lambda).
$
Now, the diagrams involved in the calculation of $\Sigma$ (at criticality) are ultraviolet convergent, so that the infinite cut-off limit can be taken. Note however that this requires that all momenta
involved in the various integrations  are small in comparison
with  $\Lambda$  or, in other words, that the integrands
are  negligibly small for momenta
$k\sim \lambda^{-1}$. Only then  can we ignore the effects of non vanishing Matsubara frequencies, and finite cut-off effects.  This implies that 
$u\lambda\sim a/\lambda$ is  sufficiently small. In this region of validity of the classical field approximation, that is, for small enough $u$, $\sigma(x,u/\Lambda)$ becomes a universal function $\sigma(x)$, independent of $u$, and   
\begin{equation}
\frac{\Delta\langle \varphi_i^2\rangle}{Nu}=-\frac{1}{2\pi ^{2}}\int dx
\frac{\sigma (x)}{x^{2}+\sigma (x)},  \label{cl-6}
\end{equation}
showing that the change in the critical density is indeed linear in $u$, and hence in $a$.

Table \ref{table} contains 
our results for $c$ 
together with some of the best estimates available in the literature.
As was the case for the critical exponents in Table \ref{tableexpo}, our numbers are all given for the PMS values $\alpha^*$ of the regulator parameter, 
and the digits quoted are those which remain stable  when 
$\alpha$ varies in the  range $[\alpha^*-\frac{1}{2},\alpha^*+\frac{1}{2}] $. 
For all $N$ values where six-loop resummed calculations exist, our results for 
$c$  are within the error bars.
For $N=100$, we find $c=2.36$, which compares well to the exact large $N$ value $c\simeq 2.33$ \cite{Baym00}.
Our  estimates for $c$ are also comparable to those obtained from 
an approximation specifically designed for this quantity \cite{Blaizot:2004qa,BMW-BE}. See also  \cite{Ledowski04}  for an earlier  estimate of $c$ using a different truncation of the RG equations, and also \cite{Floerchinger:2008jf}.

\begin{table*}[tp]
\caption{\label{table} Coefficient $c$  for the $O(N)$ models.}
\begin{ruledtabular}
\begin{tabular}{clclccll}
$N$& \multicolumn{1}{c}{BMW} &\ \  & \multicolumn{2}{c}{Resummed pert. th.}&\ \ &\multicolumn{2}{c}{Monte-Carlo} \\
    & \multicolumn{1}{c}{$c$} &
   &   \multicolumn{1}{c}{$c$} & \multicolumn{1}{c}{Ref.} &
   &   \multicolumn{1}{c}{$c$} & 
\multicolumn{1}{c}{Ref.} \\ \hline
1  &   1.15 & &    1.07(10) & \cite{Kastening03} &
 & 1.09(9) &\cite{Sun02} \\
2 &  1.37 & &    1.27(10)& \cite{Kastening03} &
   &  1.32(2) & \cite{latt2}\\
3  &    1.50 & &   1.43(11)&\cite{Kastening03} &
  &    &\\
4 &   1.63  & &    1.54(11)& \cite{Kastening03}&
    &  1.6(1)  &\cite{Sun02}\\
\end{tabular}
\end{ruledtabular}
\end{table*}

\subsection{Thermodynamics of quantum fields}

The last application that we shall consider concerns the thermodynamics of quantum fields at high temperature. This is motivated by the study of the quark-gluon plasma. In such a system one could expect a priori perturbation theory to yield accurate results because the asymptotic freedom of Quantum Chromodynamics makes the  effective coupling small at high temperature.  However, 
strict perturbation theory does not work: it exhibits indeed very poor convergence properties, even in a range of values of the coupling constant where reasonable results are obtained at $T=0$. This difference of behavior of perturbation theory at zero and finite temperature can be  understood from the  fact that, at finite temperature, the actual expansion parameter involves both the coupling constant and the magnitude of thermal fluctuations  (for a recent review, see
\cite{Blaizot:2003tw}; see also \cite{Blaizot:2009iy}).  In that respect, the problem is not specific to QCD: Similar poor convergence behavior appears also
in the simpler scalar field theory \cite{Parwani:1994zz}, and has also
been observed in the case of large-$N$ $\varphi^{4}$ theory
\cite{Drummond:1997cw}. We focus here on the case of scalar field  with a $g^2\varphi^4$ interaction  (i.e., $g^2\equiv u/24$) \cite{Blaizot:2010ut}.

Let us first recall  how 
the effect of the interactions at a given scale depends
on the magnitude of the relevant thermal fluctuations at that scale (and in some cases at a different scale as well).   The thermal fluctuations of the field are given by the following integral
\beq\label{fluctuations}
\langle \varphi^2\rangle= \int\frac{d^3 k}{(2\pi)^3}\frac{n_k}{k}, \qquad n_k=\frac{1}{{\rm e}^{k/T}-1}.
\eeq
When we perform a perturbative calculation, we assume that the ``kinetic energy'' $\sim \langle ( \del\varphi)^2\rangle$ is large compared to the ``potential energy'' $\sim g^2 \langle\varphi^4\rangle$. Obviously, this comparison depends on the strength of the coupling, but also on the typical wavelength, or momentum, of the fluctuations. To make things more precise, let us observe that the integral (\ref{fluctuations}) is dominated by the largest values of $k$ (in the absence of the statistical factor it would be quadratically divergent). One may then calculate the integral with an upper cut-off $\kappa$ and refer to the corresponding value as to ``the contribution of the fluctuations at scale $\kappa$'', and denote it by $\langle \varphi^2\rangle_\kappa$. In the same spirit, we may approximate the kinetic energy of modes at scale $\kappa$ as $ \langle ( \del\varphi)^2\rangle_\kappa \approx \kappa^2 \langle \varphi^2\rangle_\kappa$. Assuming furthermore that $ \langle\varphi^4\rangle_\kappa \approx \langle \varphi^2\rangle_\kappa^2$, one gets the expansion parameter (ratio of potential energy $\sim g^2\langle \varphi^2\rangle_\kappa^2$ to kinetic energy $\sim \kappa^2 \langle \varphi^2\rangle_\kappa$)
\beq
\gamma_\kappa=\frac{g^2 \langle \varphi^2\rangle_\kappa}{\kappa^2}.
\eeq

Let us then examine the values of this parameter for several characteristics momenta. The fluctuations that dominate the energy density at weak coupling correspond to the plasma particles and have momenta $k\sim T$. For these ``hard'' fluctuations, 
\beq\label{fluctuationsT}
\kappa\sim T,\qquad  \langle \varphi^2\rangle_T\sim T^2,\qquad \gamma_T\sim g^2.
\eeq
Thus, at this scale,  perturbation
theory works as well as at zero temperature (with expansion parameter $\sim g^2$, or rather $\alpha=g^2/4\pi$). 

The next ``natural'' scale, commonly referred to as the ``soft scale'', corresponds to $\kappa\sim gT$. We have then
\beq\label{fluctuationsgT}
\kappa\sim gT,\qquad \langle \varphi^2\rangle_{gT}\sim gT^2,\qquad \gamma_{gT}\sim g.
\eeq
(In calculating $\langle \varphi^2\rangle_\kappa$ for $\kappa\ll T$, we have used the approximation $n_k\approx T/k$, so that $\langle \varphi^2\rangle_{\kappa\ll T}\sim \kappa T$.) Since $ \gamma_{gT}\sim g$, perturbation theory can still be used to describe the self-interactions of the soft modes. However the perturbation theory is now an expansion in powers of $g$ rather than $g^2$: it is therefore less rapidly convergent. The emergence of this new expansion parameter is the origin of odd powers of $g$ in the perturbative expansion of the pressure  (such as the plasmon term $\sim g^3$). 

Another phenomenon occurs at the scale $gT$. While the expansion parameter $ \gamma_{gT}\sim g$ that controls the self-interactions of the soft fluctuations is small, the coupling between the soft modes and the thermal fluctuations at scale $T$ is not: indeed $g^2\langle \varphi^2\rangle_T\sim (gT)^2$, that is, the kinetic energy of the soft modes $\sim (gT)^2$ is comparable to their interaction energy resulting from their coupling to the hard modes, $\sim g^2\langle\varphi^2\rangle_T$. Thus the dynamics of soft modes is non-perturbatively renormalized by their coupling to hard modes. This particular coupling is encompassed by the so-called ``hard thermal loops'' \cite{HTL}. 

Finally, there is yet another scale, the ``ultra-soft scale'' $\kappa\sim  g^2 T$, at which perturbation theory completely breaks down. At this scale, we have indeed
\beq\label{fluctuationsg2T}
\kappa\sim g^2T,\qquad \langle \varphi^2\rangle_{gT}\sim g^2T^2,\qquad \gamma_{g^2T}\sim 1.
\eeq
Thus the ultra-soft  fluctuations remain strongly coupled for arbitrarily small values of the coupling constant. This situation does not  occur for a scalar field since a mass is generated at scale $gT$, which renders the contribution of the fluctuations at the scale $g^2T$ negligible.  However this situation is met in QCD for the long wavelength, unscreened, magnetic
fluctuations. 

These considerations suggest that  the  main difficulty with thermal perturbation
theory is 
not so much related to the magnitude of the coupling constant (for the relevant temperatures it is not that large), but it is rather due 
 to the interplay of degrees of freedom with various
wavelengths, possibly involving collective modes.  In some sense, field theories at high temperature are multiscale systems. At weak coupling the dynamically generated scales $T$, $gT$ and $g^2 T$ are well separated. This  allows for instance the organization of the calculation using effective field theory. However the scale separation disappears when  the coupling is not too small: then, the various degrees of freedom mix and the situation requires a different type of analysis. 

The exact renormalization group is ideally suited to cope with this type of situations. In particular, as we have seen,  the BMW truncation scheme provides an excellent description of the momentum dependence of the 2-point function, from the low momenta of the critical region, all the way up to the large momenta of the perturbative regime. One may then expect this method to capture accurately the contributions to the thermodynamical functions  of thermal fluctuations  from various momentum ranges, and hence handle properly the mixing between degrees of freedom that takes place as the coupling grows. Since it involves also non trivial momentum dependent self-energies, the method  also encompasses implicitly effects related to the damping of quasiparticles, or their coupling to complex multi-particle configurations. 

Fig.~\ref{fig:masspressure} displays the pressure as the function of the coupling constant at the scale $g(2\pi T)$. The various diverging curves labelled $g^2,\cdots, g^8\ln g$ indicate the results of perturbative calculations, up to order $g^8\ln g$ (for recent high order calculations of the thermodynamics of the scalar field, see \cite{Andersen:2009ct}). These curves clearly illustrate the poor behavior of strict perturbation theory. The other curves correspond to various implementations of the exact renormalization group, as well as to a  2PI calculation based on a simple 2-loop skeleton \cite{Blaizot:2006rj}. The two curves labelled LPA correspond to two different choices of regulators: either the Litim regulator \cite{Litim}, which is implemented only for three-momenta, with the (untruncated) sums over the Matsubara frequencies being performed analytically) \cite{Blaizot:2006rj}, and an exponential regulator that affects both the momenta and the frequencies. The BMW approximation scheme is better justified  when one uses an Euclidean symmetric four dimensional regulator, and the calculations reported here have been done with the exponential regulator (\ref{regulator}). 

\begin{figure}
\hfill{}\includegraphics[scale=1.2]{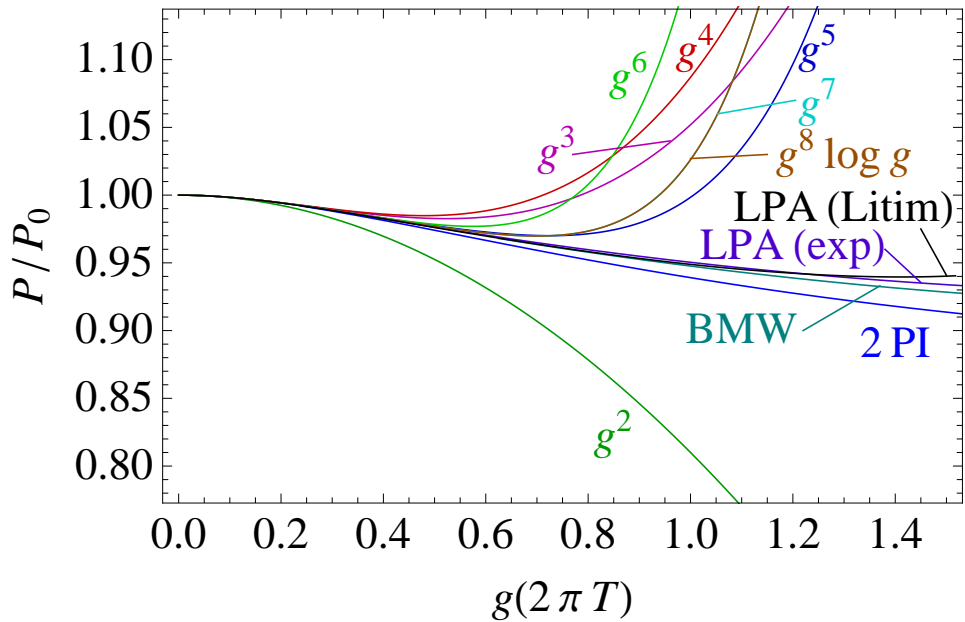}\hfill{}

\caption{({\it Color online}) Pressure as function of the coupling. The various resummation and RG methods compared are 2 PI, LPA with exponential regulator (exp), LPA with Litim regulator, and BMW. Shown are also perturbative results through order $g^6 \log g$ for the mass and $g^8 \log g$ for the pressure. The $g^7$ and $g^8 \log g$ curves for the pressure almost lie on top of each other, as do the BMW and LPA (exp) curves for the mass.\label{fig:masspressure}. From Ref.~\cite{Blaizot:2010ut}}

\end{figure}

 In contrast to the perturbative calculation, the  calculations based on the renormalization group show a remarkable stability, and a smooth extrapolation towards strong coupling. 
As it turns out, the results obtained are not too different from those of the LPA, nor from the simple 2-loop 2PI approximation  used in Ref.~\cite{Blaizot:2006rj}. In physical terms, both the LPA and the 2-loop 2PI approximation  correspond to approximations where the degrees of freedom of the hot scalar plasma are massive quasiparticles. The new scheme goes beyond that simple picture. 
This stability of the results against improvements in the approximation suggests that the scheme that we are using to solve the  NPRG equations may give already, at the level at which it is implemented here, an accurate representation of the exact pressure, and this over a wide range of coupling constants. It also indicates that for such a system the quasiparticle picture is presumably robust.

\section{Conclusion}

The few applications that are presented in the previous section illustrate the power of the renormalization group, when coupled to an approximation scheme that allows for a determination of the full momentum dependence of the $n$-point functions. The equations that need to be solved  are a priori complicated: these are flow equations which are at the same time partial differential equations (with partial derivatives with respect to the background field), with integral kernels that involve the solution itself. Still they can be solved at a rather modest numerical cost, using elementary numerical techniques. 
The studies presented here involved only the leading-order of a 
systematic approximation scheme. In the absence of any small parameter controlling the magnitude of successive orders, a study of the next order would be necessary in order to quantify the accuracy that has been reached. However,  the robustness of the leading order results can already be gauged from their weak residual 
dependence on the choice of the regulator.\\

\noindent{\bf Acknowledgements} Most of the work summarized here results from much enjoyable  collaborations with F. Benitez, H. Chat\'e, B. Delamotte, A. Ipp, T. Matsuura, R. Mendez Galain, J. Pawlowski, U. Reinosa, and N. Wschebor, whose respective contributions to the various topics discussed here can be easily inferred form the references given. I would like to thank also the organizers for their invitation to the 2010 INT workshop on ``New applications of the renormalization group method'', and especially Y. Meurice and S.-W. Tsai for their persevering encouragements to complete this write up.

\end{document}